\documentclass[a4paper,twocolumns,11pt,accepted=2024-03-04]{quantumarticle}
\pdfoutput=1
\usepackage[utf8]{inputenc}
\usepackage[english]{babel}
\usepackage[T1]{fontenc}
\usepackage{amsmath, amssymb}
\usepackage{hyperref}

\usepackage{cite}

\usepackage{verbatim}
\usepackage{physics}
\usepackage[numbers]{natbib}
\usepackage{enumitem}
\usepackage{booktabs}

\usepackage[colorlinks=true]{hyperref}
\hypersetup{citecolor = blue, linkcolor=blue, urlcolor=blue}

\newcommand{\shadownorm}[1]{\ensuremath{\| #1 \|_{\rm sh}}}
\newcommand{\CZ}{\ensuremath{\textsf{CZ}}}
\newcommand{\CPhase}[1]{\ensuremath{\textsf{CPhase}(#1)}}

\begin{document}

\title{Classical shadows based on locally-entangled measurements}

\author{Matteo Ippoliti}
\affiliation{Department of Physics, The University of Texas at Austin, Austin, TX 78712, USA}
\affiliation{Department of Physics, Stanford University, Stanford, CA 94305, USA}

\begin{abstract}
We study classical shadows protocols based on randomized measurements in $n$-qubit entangled bases, generalizing the random Pauli measurement protocol ($n = 1$). 
We show that entangled measurements ($n\geq 2$) enable nontrivial and potentially advantageous trade-offs in the sample complexity of learning Pauli expectation values.
This is sharply illustrated by shadows based on two-qubit Bell measurements: the scaling of sample complexity with Pauli weight $k$  improves quadratically (from $\sim 3^k$ down to $\sim 3^{k/2}$) for many operators, while others become impossible to learn.
Tuning the amount of entanglement in the measurement bases defines a family of protocols that interpolate between Pauli and Bell shadows, retaining some of the benefits of both.
For large $n$, we show that randomized measurements in $n$-qubit GHZ bases further improve the best scaling to $\sim (3/2)^k$, albeit on an increasingly restricted set of operators.
Despite their simplicity and lower hardware requirements, these protocols can match or outperform recently-introduced ``shallow shadows'' in some practically-relevant Pauli estimation tasks.
\end{abstract}

\maketitle

\section{Introduction}

Classical shadows are a powerful method to learn many properties of unknown quantum states with a relatively low number of measurements~\cite{huang_predicting_2020,elben_randomized_2023,hadfield_measurements_2020,chen_robust_2021,acharya_shadow_2021,struchalin_experimental_2021,levy_classical_2021,kunjummen_shadow_2023,huang_learning_2022-1,wan_matchgate_2022,bu_classical_2022,nguyen_optimizing_2022,koh_classical_2022,grier_sample-optimal_2022,becker_classical_2022,seif_shadow_2023,van_kirk_hardware-efficient_2022}. 
This is an important task in light of the advent of programmable quantum simulators capable of preparing increasingly complex quantum states, whose experimental characterization and classical description may be challenging~\cite{arute_quantum_2019,altman_quantum_2021,ebadi_quantum_2021,mi_information_2021}. 
Classical shadows are based on randomized measurements~\cite{brydges_probing_2019,elben_statistical_2019,elben_randomized_2023}: the unknown state of interest is measured in a large number of different bases, randomly chosen from a suitable ensemble, and the resulting classical data is stored and processed to predict properties of the state. 

Different choices for the ensemble of random unitary rotations yield different flavors of classical shadows, each of which may be best suited to the prediction of different properties~\cite{huang_predicting_2020}.
Arguably the most practically-relevant example is the {\it random Pauli} ensemble, where each qubit is measured in a randomly-chosen $X$, $Y$ or $Z$ basis; this requires only single-qubit random rotations on the hardware and is well suited to learning e.g. the expectation value of $k$-local operators. 
Another important example is the {\it random Clifford} ensemble, where the basis is randomized by a global Clifford operation; this allows efficient estimation of fidelities and low-rank operators. 
Intermediate schemes dubbed {\it shallow shadows} have been recently introduced~\cite{akhtar_scalable_2023,bertoni_shallow_2022,arienzo_closed-form_2022,ippoliti_operator_2023}. These randomize the basis by means of variable-depth circuits, thus interpolating between random Pauli and random Clifford measurements.

In this work, we introduce a family of protocols that interpolates between locally- and globally-random measurements in a different way, by tuning the locality of subsystems on which entangled measurement bases are allowed. 
The protocols are hardware-efficient, requiring only few-body entanglement, and the classical post-processing is likewise simple. Nonetheless, they can outperform random Pauli shadows and even shallow shadows in some Pauli estimation tasks of practical interest, making them a useful addition to the randomized measurement toolbox.

\subsection{Review}

In general, given an ensemble of random unitaries, the classical shadows protocol is as follows. A quantum state of interest $\rho$ on $N$ qubits is transformed under a unitary $U$ drawn from the ensemble. It is then measured in the computational basis yielding a bitstring $b$. The pairs $\{ (U, b) \}$ of basis choice and measurement outcome represent classical data that can be used to efficiently construct a compressed description of the state. Namely one builds ``snapshots'' $\hat{\sigma} = U^\dagger \ketbra{b}{b} U$ which in expectation are related to the state of interest $\rho$ by a channel $\mathcal{M}$, called the {\it shadow channel}: $\mathbb{E}[\hat\sigma] = \mathcal{M}(\rho)$.
From this, one defines ``inverted snapshots'' $\hat{\rho} = \mathcal{M}^{-1}(\hat{\sigma})$ which by construction yield $\rho$ in expectation~\cite{huang_predicting_2020}. The set of inverted snapshots $\{\hat{\rho}_i\}_{i=1}^K$ (obtained over $K$ iterations fo the quantum experiment) is a {\it classical shadow} of $\rho$ of size $K$.

The practical usefulness of classical shadows depends on the {sample complexity} of various estimation tasks, i.e., how many experimental shots are needed in order to predict a given property of $\rho$ to a fixed (additive) accuracy $\epsilon$ with high probability.
For predicting $M$ linear functions (i.e. expectation values of a set of operators $\{O_i\}_{i=1}^M$), this is bounded above by~\cite{huang_predicting_2020}
\begin{equation}
\log(M) \frac{\max_i\shadownorm{O_i}^2}{\epsilon^2},
\label{eq:hkp_bound}
\end{equation}
where $\shadownorm{\cdot}$ is a norm determined by the measurement protocol, called the {\it shadow norm}.
For a Pauli operator $P$ in an $N$-qubit system, assuming Pauli invariance of the measurement ensemble~\cite{bu_classical_2022}, the shadow norm is given by 
\begin{equation}
    \shadownorm{P}^2 = \frac{1}{2^N} \Tr(P \mathcal{M}^{-1}(P)),
\end{equation}
independent of the state $\rho$. Here the inverse map $\mathcal{M}^{-1}$ appears due to the construction of the ``inverted snapshots'' $\hat{\rho} = \mathcal{M}^{-1}(\hat{\sigma})$. 
For random Pauli measurements, the shadow channel factors into single-qubit depolarizing channels $\mathcal{M} = \mathcal{E}_1^{\otimes N}$, with $\mathcal{E}_1(\sigma^\alpha) = \lambda_\alpha \sigma^\alpha$, $\alpha \in \{0,x,y,z\}$, and eigenvalues
\begin{equation}
    \lambda_{\circ} = 1,
    \qquad 
    \lambda_\bullet = \frac{1}{3}, \label{eq:evals_paulishadows}
\end{equation}
where $\circ$ denotes the identity ($\alpha = 0$) and $\bullet$ stands for any of the traceless Pauli matrices $\alpha \in \{x,y,z\}$.
It follows that 
\begin{equation}
\shadownorm{P}^2 = \lambda_\circ^{k-N} \lambda_\bullet^{-k} = 3^k,
\label{eq:randompauli_scaling}
\end{equation}
where $k$ is the {\it weight} of operator $P$, i.e., the number of qubits on which it acts nontrivially~\cite{huang_predicting_2020}. 

A simple way to understand this scaling is that there are three possible basis choices ($X$, $Y$ or $Z$) per site, sampled randomly in each experimental run. Only some basis choices are useful towards the estimation of a Pauli expectation value $\langle P \rangle$. 
In particular, only measurement bases that correctly match {\it all} nontrivial Pauli matrices in $P$ contribute to its estimation (in the language of Ref.~\cite{huang_efficient_2021}, we say such measurements ``hit'' $P$). As only one in $3^k$ bases ``hits'' $P$, estimating $\langle P \rangle$ to accuracy $\epsilon$ requires of order $3^k /\epsilon^2$ iterations of the experiment.

It is known on information-theoretic grounds~\cite{huang_predicting_2020} that the scaling of $3^k$ is optimal in general. However, it is possible to improve performance on certain Pauli operators at the expense of others. 
For example, shallow shadows~\cite{akhtar_scalable_2023,bertoni_shallow_2022} were shown to achieve a scaling of $\sim k 2^k$ (when tuned to a $k$-dependent optimal depth) for Pauli operators with contiguous support in one dimension~\cite{ippoliti_operator_2023}, while typically performing worse on operators with a sparse support.
Such trade-offs may be worthwhile in many cases, given the importance of geometric locality in many-body physics. 
The protocols we introduce in this work feature a similar trade-off, with very favorable performance on a physically-relevant class of operators obtained at the expense of the learnability of other operators. 

\begin{figure}
    \centering
    \includegraphics[width=\columnwidth]{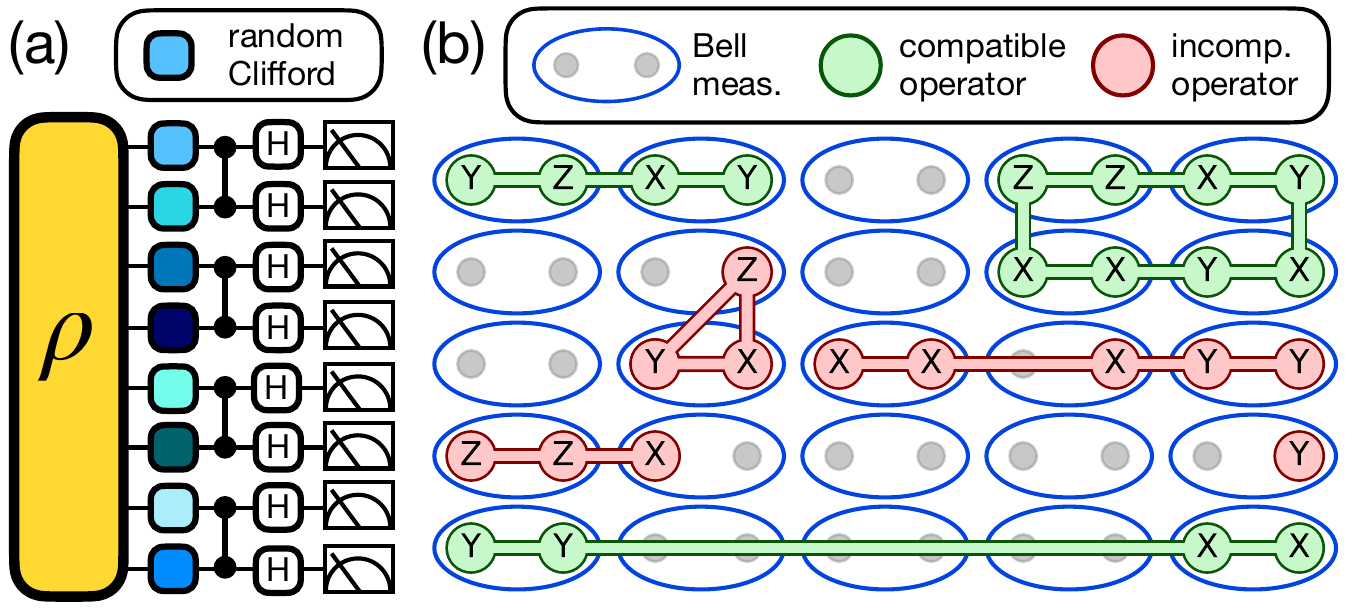}
    \caption{Schematic of the Bell shadows protocol. 
    (a) Unknown state $\rho$ is locally scrambled with random single-qubit Clifford gates (colored squares), then pairs of neighboring qubits are measured in the Bell basis (shown as a sequence of $\CZ$, Hadamard and computational basis measurements).
    (b) Bell measurements define a dimer covering of the lattice (ellipses). An operator is compatible with the covering if its support intersects each dimer on 0 or 2 sites (green), incompatible otherwise (red). Bell shadows only learn compatible operators.}
    \label{fig:schematic}
\end{figure}

\section{Bell shadows} \label{sec:bell}

\subsection{Protocol} \label{sec:bell_setup}

We begin by introducing a variant of the random Pauli measurement protocol based on two-qubit measurements in the Bell basis. 
The protocol requires first to choose a grouping of the qubits into pairs (we assume the number of qubits $N$ is even). For concreteness, we take this pairing to be geometrically-local on an underlying lattice, and thus refer to this as a {\it dimer covering} of the lattice; however geometric locality is not necessary. 
The protocol then involves the following steps, illustrated in Fig.~\ref{fig:schematic}(a):
\begin{enumerate}[label=(\roman*)]
\item ``Locally scramble'' the state, $\rho \mapsto U\rho U^\dagger$ with $U = \bigotimes_{i=1}^N u_i$, each $u_i$ being a random Clifford gate; 
\item Measure each pair of qubits in the Bell basis;
\item Build the classical shadow according to the standard prescription~\cite{huang_predicting_2020}. 
\end{enumerate}

In Fig.~\ref{fig:schematic}(a), the Bell measurement [step (ii)] is compiled as a sequence of $\CZ$, Hadamard and $X$-basis measurements.
$\CZ$ is the controlled-$Z$ gate, $\CZ = \textsf{diag}(1,1,1,-1)$; it is a Clifford operation that maps the Pauli-$X$ basis to the basis stabilized by $\pm X_1 Z_2$, $\pm Z_1 X_2$. 
This basis, up to local unitary transformations (which may be absorbed into the local scrambling step), is equivalent to the standard Bell basis stabilized by $\pm X_1 X_2$, $\pm Z_1 Z_2$, i.e. the basis vectors $\{\ket{\Phi^\alpha}: \alpha = 0,x,y,z\}$, with $\ket{\Phi^0} = \frac{1}{\sqrt{2}}(\ket{00} + \ket{11})$ and $\ket{\Phi^\alpha} = (\sigma^\alpha\otimes I)\ket{\Phi^0}$.

\begin{figure*}
    \centering 
    \includegraphics[width=0.8\textwidth]{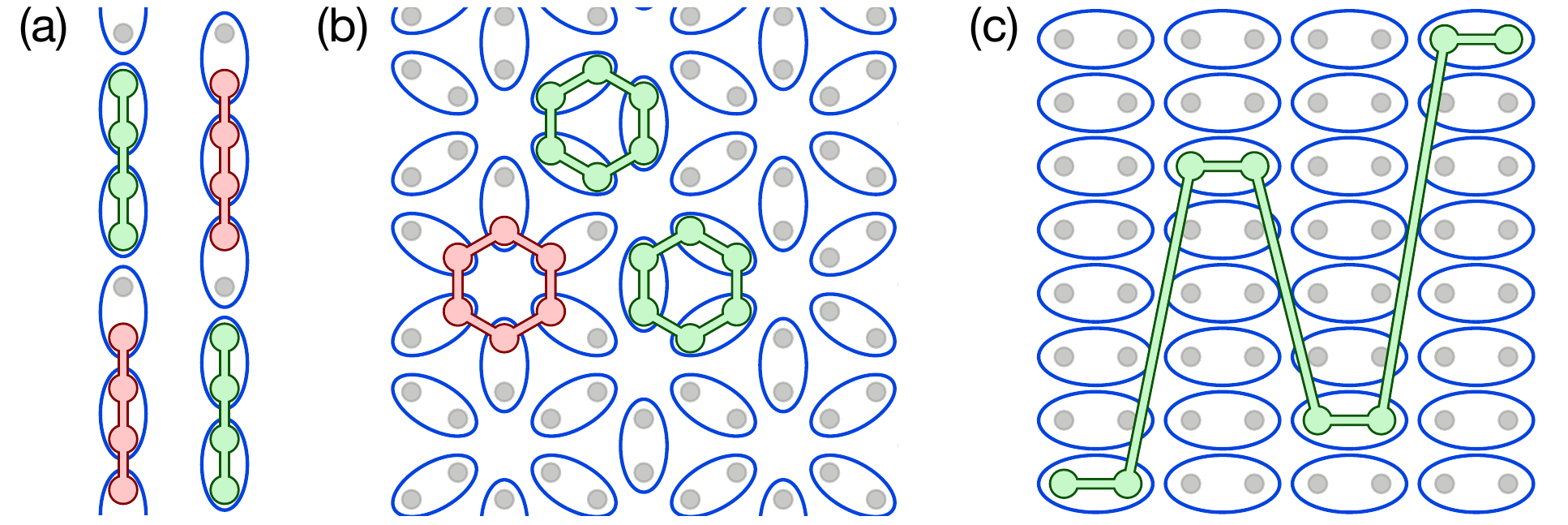}
    \caption{Use cases of Bell shadows, Sec.~\ref{sec:bell_app}.
    The graphical conventions are as in Fig.~\ref{fig:schematic}(b): blue ellipses denote dimers whose qubits are measured in a Bell basis, green operators are compatible with the given dimer covering and thus learnable, red operators are incompatible and not learnable.
    (a) Pauli string operators in 1D chains. All operators of even length are learnable by sampling two distinct dimer coverings of the chain (left and right). 
    (b) Plaquette operators of a honeycomb system (e.g. color code stabilizers). The dimer covering in the picture is compatible with two thirds of all hexagonal plaquettes; the remaining ones may be learned by translating the dimer covering by one lattice vector. 
    (c) Multi-point functions of two-body operators.
    \label{fig:applications}}
\end{figure*}

\subsection{Sample complexity}\label{sec:bell_sc}

The sample complexity of learning the expectation value of a Pauli operator $P$ via this protocol is determined by the shadow norm $\shadownorm{P}^2$.
This can be analyzed in analogy with the random Pauli measurement protocol~\cite{huang_predicting_2020}, by noting that the shadow channel factors into a product of two-qubit channels, $\mathcal{M} = \mathcal{E}_2^{\otimes N/2}$.
Each two-qubit channel $\mathcal{E}_2$ acts on a single dimer as
\begin{align}
    \mathcal{E}_2(\rho) 
    & = \sum_\alpha \int{\rm d}u\, {\rm d}v\, 
    \bra{\Phi^\alpha}u\otimes v \rho (u \otimes v)^\dagger \ket{\Phi^\alpha} \nonumber \\
    & \qquad \times (u \otimes v)^\dagger \ket{\Phi^\alpha}\bra{\Phi^\alpha} u\otimes v.
\end{align}
Here both ${\rm d}u$ and ${\rm d}v$ are the Haar measure\footnote{While the protocol is defined in terms of random Clifford gates, we may use the 2-design property to replace the sum over Clifford gates with the integral over the Haar measure.} over $U(2)$. 
The channel is diagonal in the Pauli basis owing to local scrambling~\cite{akhtar_scalable_2023,bertoni_shallow_2022}, $\mathcal{E}_2(\sigma^\alpha\otimes \sigma^\beta) = \lambda_{\alpha\beta} \sigma^\alpha \otimes \sigma^\beta$. The eigenvalues are given by
\begin{align}
    \lambda_{\alpha \beta}
    & = \frac{1}{4} \Tr[\sigma^\alpha \otimes \sigma^\beta \mathcal{E}_2(\sigma^\alpha\otimes \sigma^\beta)] \nonumber \\
    & = \int{\rm d}u\ {\rm d}v\ \bra{\Phi^0}(u\sigma^\alpha u^\dagger) \otimes (v\sigma^\beta v^\dagger) \ket{\Phi^0}^2. \label{eq:lambda_matrixelement}
\end{align}
Note that the four possible Bell state outcomes $\{\ket{\Phi^\gamma}:\gamma=0,x,y,z\}$ all yield the same contribution to Eq.~\eqref{eq:lambda_matrixelement} due to unitary invariance of the ${\rm d}u$ measure, hence a factor of 4.

Next, using the fact that $\bra{\Phi^0} B\otimes C\ket{\Phi^0} = \frac{1}{2}\sum_{i,j} B_{ij} C_{ij} = \frac{1}{2} \Tr(BC^T)$ for any single-qubit operators $B$, $C$, we have
\begin{equation}
    \lambda_{\alpha \beta} 
    = \frac{1}{4} \int{\rm d}u\ {\rm d}v\ \left[\Tr(u\sigma^\alpha u^\dagger v \sigma^\beta v^\dagger) \right]^2 \label{eq:lambda_integral}
\end{equation}
We see that one of the random rotations is redundant: 
since the integrand depends on $u$, $v$ only through the product $u^\dagger v$, we may change integration variables from $\int {\rm d}u\int {\rm d}v$ to $\int {\rm d}u\int {\rm d}(u^\dagger v)$ (using the unitary invariance of the Haar measure) and integrate over $u^\dagger v$; the result is independent of $u$, making it redundant.
%e.g., the rotation $v$ may be absorbed into the (unitarily invariant) measure ${\rm d}u$, or vice versa.
This illustrates an advantageous property of the protocol: it is enough to scramble only one qubit per dimer.
Physically, this is a consequence of gate teleportation across the Bell pair: $(u\otimes I)\ket{\Phi^0} = (I\otimes u^T)\ket{\Phi^0}$.

From Eq.~\eqref{eq:lambda_integral}, it is straightforward to derive the results
\begin{equation}
    \lambda_{\circ\circ} = 1,\quad
    \lambda_{\circ\bullet} = \lambda_{\bullet\circ} = 0, \quad 
    \lambda_{\bullet\bullet} = \frac{1}{3}
    \label{eq:evals1}
\end{equation}
(like in Eq.~\eqref{eq:evals_paulishadows}, $\circ$ stands for the identity, $\alpha = 0$, while $\bullet$ stands for a traceless Pauli matrix $\alpha \in \{x,y,z\}$).
These eigenvalues fully determine the shadow norm of any Pauli operator $P$: 
\begin{equation}
    \shadownorm{P}^2 = \prod_{(i,j)} \lambda_{\alpha_i \alpha_j}^{-1}
\end{equation}
where the pair $(i,j)$ ranges over dimers in the system and $\alpha_i$ indices are defined by $P = \bigotimes_{i=1}^N \sigma^{\alpha_i}_i$.

The presence of null eigenvalues in $\mathcal{M}$ immediately shows that the ensemble is not tomographically complete.
For convenience, let us introduce the following definition: a Pauli operator $P$ is \emph{compatible} with the dimer covering if its support intersects each dimer in either 0 or 2 sites.
Clearly Pauli operators that are incompatible with the dimer covering are not learnable, as sketched in Fig.~\ref{fig:schematic}(b), since they feature at least one $\lambda_{\circ\bullet} = 0$ eigenvalue.
Nonetheless, for the Pauli operators that are compatible with the dimer covering, the shadow norm is remarkably low: 
\begin{equation}
    \shadownorm{P}^2 = \lambda_{\circ\circ}^{k/2-N} \lambda_{\bullet\bullet}^{-k/2} = 
    3^{k/2} \label{eq:shdn}
\end{equation}
where $k$ is the weight of $P$. 
There are $10^{N/2}$ compatible Pauli operators out of a total of $4^N$ (each one of the $N/2$ dimers may host $\sigma^0 \otimes \sigma^0$ or $\sigma^\alpha\otimes \sigma^\beta$ with $\alpha,\beta\in \{x,y,z\}$, for a total of 10 options). This exponentially vanishing fraction, $\simeq 0.79^N$, is the price to pay for the improved sample complexity. 

In analogy with the random Pauli case, we may understand the performance of Bell shadows with a simple basis counting argument. For each qubit pair, each run of the experiment measures 3 out of the 9 two-qubit operators $\{\sigma^\alpha\otimes \sigma^\beta:\ \alpha,\beta = x, y, z\}$---e.g., the experimentalist may explicitly measure $XX$ and $YZ$, but that also implicitly measures\footnote{As an example, let us imagine measuring $XX = +1$ and $YZ = -1$; this also informs us of the eigenvalue of $ZY = XX \cdot YZ$: in this case, $(+1)\cdot(-1) = -1$. More generally, measuring any number of commuting Pauli operators $\{P_i\}$ is equivalent to measuring the entire {\it stabilizer group} they generate.} $ZY = XX\cdot YZ$. Thus we measure 3 out of the 9 weight-two operators on the two qubits. In all, the probability that the measurement ``hits'' a given $k$-qubit Pauli operator $P$ (compatible with the dimer covering) is $3/9 = 1/3$ per {\it pair} of qubits on which $P$ acts nontrivially, i.e. $3^{-k/2}$ overall. We also see why operators that are incompatible with the dimer covering are not learnable:
any triplet of weight-2 commuting operators has the structure of $\{X\otimes \sigma^\alpha, Y\otimes \sigma^\beta, Z\otimes \sigma^\gamma\}$, with $(\alpha \beta \gamma)$ a permutation of $(xyz)$; in other words, each operator in the triplet must have a different Pauli matrix on the first qubit. The same clearly goes for the second qubit. Thus weight-1 operators such as $IX$ or $YI$ necessarily anticommute with two of the operators in the triplet, and therefore we never learn about their value from Bell measurements.

\subsection{Some use cases}\label{sec:bell_app}

While tomographically incomplete, the random Bell measurement ensemble is very powerful for learning Pauli operators compatible with the chosen dimer covering. 
This includes many cases of interest for condensed matter physics and quantum information science; we list some examples below.

\paragraph{String operators.} In a 1D lattice, any Pauli operator $P$ whose support is made of $k$ consecutive sites, with $k$ even, is learnable. Operators of this form are interesting in condensed matter physics as they include string order parameters for symmetry-protected topological (SPT) phases~\cite{haegeman_order_2012}. To learn {\it all} operators of this form with a given length $k$, regardless of endpoint location, one must sample two dimer coverings of the lattice (pairing even and odd bonds, respectively, see Fig.~\ref{fig:applications}(a)). Following Eq.~\eqref{eq:hkp_bound},
the overall sample complexity of learning {all} $M = N\cdot 3^k$ operators of this form ($N$ possible endpoint locations, $3^k$ sequences of Pauli matrices inside the support) with accuracy $\epsilon$ is at most $2 \ln(M/2) 3^{k/2}\epsilon^{-2}$. 
Here we have split the $M$ operators into two sets of size $M/2$ based on the parity of their endpoint location, which determines the choice of dimer covering, and applied Eq.~\eqref{eq:hkp_bound} to each set separately.
The corresponding scaling for random Pauli measurements is $\ln(M) 3^k \epsilon^{-2}$, larger by a factor of $\simeq 3^{k/2} / 2$. 

\paragraph{Plaquette operators.}
Products of Pauli matrices around a plaquette in a two-dimensional lattice may also be learnable by choosing suitable dimer coverings.
Stabilizers of topological codes~\cite{bombin_introduction_2013} typically take this form, as do ``ring exchange'' terms~\cite{thouless_exchange_1965} in lattice Hamiltonians.
As an example, all hexagonal plaquette operators of a honeycomb qubit lattice may be learned by sampling two dimer coverings (Fig.~\ref{fig:applications}(b)). The prefactor to the sample complexity is $2\cdot 3^3 = 54$ (i.e. the squared shadow norm $3^{k/2}$, $k=6$, for each of the 2 dimer coverings). 
For comparison, with random Pauli measurements the corresponding prefactor is $3^6 = 729$, over an order of magnitude larger.

\paragraph{Multi-point functions.}
Learnable operators need not be geometrically local; 
in particular, multi-point correlation functions of $k$-body Hamiltonian terms (with $k$ even) may be learnable. These quantities play an important role in condensed matter physics, both in and out of equilibrium; for example, two-point functions may capture long-range order or describe the propagation of a single quasiparticle from one place to another, while four-point functions may describe {scattering} between two quasiparticles, with implications for pairing instabilities or transport~\cite{altland_condensed_2010,chowdhury_multipoint_2013,takacs_correlation_2018,kugler_multipoint_2021}. 
As a simple example, consider a local Hamiltonian in a $d$-dimensional lattice, $H = \sum_{\langle i , j \rangle} h_{(i,j)}$, where $h_{(i,j)} = \sum_{\alpha,\beta = x,y,z} J_{i,j}^{\alpha\beta}\sigma_i^\alpha \sigma_j^\beta$ is the energy density operator at bond $b = (i,j)$, and $J_{i,j}^{\alpha\beta}$ are {\it a priori} unknown couplings\footnote{With some prior knowledge, one could further improve performance by biasing or derandomization, which we ignore here.}.
Bell shadows can efficiently estimate $p$-point functions $\langle h_{b_1} h_{b_2}\cdots h_{b_p}\rangle$ of the energy density (as long as all bonds $b_1,\dots b_p$ match the dimer covering, see Fig.~\ref{fig:applications}(c)) with sample complexity scaling with $p$ as $\sim 3^p$, compared to the scaling for random Pauli measurements $\sim 9^{p}$.

\section{General two-qubit bases} \label{sec:quasi-bell}

Despite the applicability of Bell shadows to several problems of interest, it would be desirable to have a {\it tomographically complete} ensemble with similar properties.
To this end, one can consider general two-qubit entangled bases, which may interpolate between random Pauli and random Bell measurements.

For a general locally-scrambled measurement ensemble~\cite{hu_classical_2023,akhtar_scalable_2023}, the eigenvalues of the shadow channel are fully specified by the 
{\it entanglement feature}~\cite{you_entanglement_2018,kuo_markovian_2020} $\{\overline{\mathcal P}_A\}$ of the measurement basis: a vector of length $2^N$, with entries labelled by subsystems $A \subseteq \{1,\dots N\}$, defined by
\begin{equation}
\overline{\mathcal P}_A \equiv \mathbb{E}_\psi {\rm Tr}[ ({\rm Tr}_{\bar A} \ketbra{\psi})^2],
\label{eq:ent_feature}
\end{equation}
where $\ket{\psi}$ ranges over states in the measurement basis (we assume the measurements are projective and thus described by a basis of pure states). In words, Eq.~\eqref{eq:ent_feature} describes the purity of each subsystem $A$ averaged over basis states.
As shown in Ref.~\cite{bu_classical_2022}\footnote{See in particular Proposition 9 in Ref.~\cite{bu_classical_2022}, which is equivalent to Eq.~\eqref{eq:lambda_A} up to notation. A derivation with the same notation used here is also given in Ref.~\cite{ippoliti_learnability_2023}, Section V.A.}, the shadow channel eigenvalues are given by
\begin{equation}
    \lambda_A = (-1/3)^{|A|} \sum_{B\subseteq A} (-2)^{|B|} \overline{\mathcal P}_B \label{eq:lambda_A}
\end{equation}
(here $\lambda_A$ refers to any Pauli operator with support $A$).
In the present (two-qubit) case, the entanglement feature contains only one nontrivial parameter: the average purity of a single qubit, which we will denote by\footnote{The `$a$' in $S_2^a$ stands for {\it annealed}, since the definition $S_2^a \equiv -\log \mathbb{E}_\rho {\rm Tr}(\rho^2)$ is analogous to an ``annealed average'' in statistical mechanics, as opposed to a ``quenched average'' $\overline{S_2} = -\mathbb{E}_\rho \log {\rm Tr}(\rho^2)$. The two differ by the order of averaging and taking the logarithm. We have $S_2^a \leq \overline{S_2}$ by convexity. } $e^{-S_2^a}$.
In practice this parameter can be tuned by varying the two-qubit gates in the protocol, Fig.~\ref{fig:schematic}(a), from $\CZ$ to $\CPhase{\phi}$; the gate angle $\phi$ is related to the entropy $S_2^a$ via $e^{-S_2^a} = \cos^4(\phi/4) + \sin^4(\phi/4)$, which achieves the extremal values at $\phi = 0$ ($S_2^a = 0$, disentangled measurements) and $\phi = \pi$ ($S_2^a = \ln(2)$, Bell measurements). 

It is convenient to introduce a ``deformation parameter'' $\delta = \ln(2) - S_2^a$, so that $\delta = 0$ recovers Bell shadows. Then, from Eq.~\eqref{eq:lambda_A}, the nontrivial eigenvalues of the two-qubit shadow channel $\mathcal{E}_2$ are
\begin{equation}
    \left\{ 
    \begin{aligned}
    & \lambda_{\bullet\circ} =
    \lambda_{\circ \bullet} = 
    \frac{e^\delta-1}{3}\simeq \frac{\delta}{3}, \\
    & \lambda_{\bullet \bullet} = \frac{5-2e^\delta}{9} \simeq \frac{1}{3+2\delta},
    \end{aligned} 
    \right.
    \label{eq:evals2}
\end{equation}
where the $\simeq$ holds to leading order in $\delta \to 0$. 

Thus for $\delta >0$ the spectrum of $\mathcal{M}$ becomes strictly positive, and the ensemble tomographically complete.
The shadow norm of a Pauli operator $P$ with support $A$ that cuts $c_A$ dimers is given by 
\begin{align} 
    \shadownorm{P}^2 
    & = \lambda_{\bullet\bullet}^{-(|A|-c_A)/2} \lambda_{\bullet\circ}^{-c_A} \label{eq:shdn_delta} \\
    & \simeq (3+2\delta)^{|A|/2} \left(\frac{\sqrt{3}}{\delta} \right)^{c_A}
    \label{eq:shdn_delta_approx}
\end{align}
where the second line is up to subleading corrections in small $\delta$.
We recover Eq.~\eqref{eq:shdn} for $\delta \to 0$: operators that are compatible with the dimer covering ($c_A=0$) have squared shadow norm $3^{|A|/2}$, the others ($c_A>0$) are not learnable.

\begin{figure}
    \centering
    \includegraphics[width=\columnwidth]{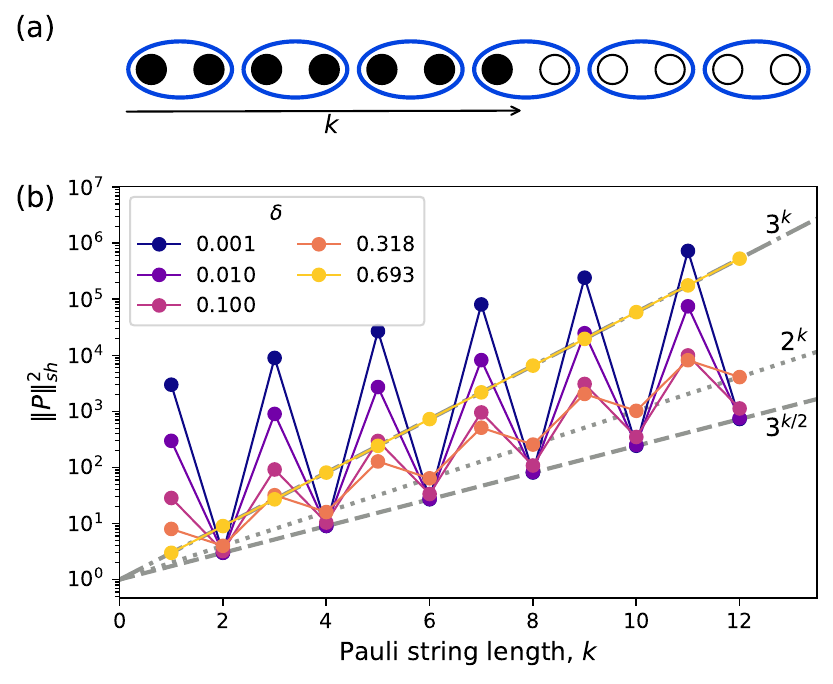}
    \caption{Pauli shadow norm from two-qubit measurements in bases with variable entanglement. 
    (a) Setup: qubits (circles) in a 1D chain are grouped into pairs (blue ellipses) and measured in a given two-qubit basis. The Pauli operator's support (full circles) has length $k$. Operators with odd $k$ break a dimer.
    (b) $\shadownorm{P}^2$ as a function of $k$ for different values of the basis entanglement $S_2 = \ln(2) - \delta$. 
    Also shown are the scalings $3^{k/2}$ (dashed line), $2^k$ (dotted line) and $3^k$ (dot-dashed line). \label{fig:qBshadows}}
\end{figure}

For the practically-relevant case of string operators in 1D, the result Eq.~\eqref{eq:shdn_delta} is illustrated in Fig.~\ref{fig:qBshadows}.
For small $\delta$ (highly entangled measurement basis), the asymptotic scaling in large $k$ is close to $3^{k/2}$, but odd-$k$ operators (which break a dimer) are much more costly to learn.
Increasing $\delta$ alleviates the even-odd discrepancy at the expense of a slightly worse asymptotic scaling in $k$.
In particular for $\delta = \ln(11/8) \simeq 0.318$, we have $\shadownorm{P}^2 = 4^{k\, {\rm mod}\, 2} \cdot 2^k$.
This is notable as it beats the performance of optimal-depth shallow shadows ($\sim k 2^k$)~\cite{ippoliti_operator_2023} for modest values $k\gtrsim 4$ (including odd $k$) that are relevant to near-term applications. It is especially remarkable given the much simpler protocol; see Sec.~\ref{sec:previous} for further discussion of the relationship between these results.

Finally, it is interesting to relax the condition of a fully-contiguous support to allow for a density of ``holes'' (identity matrices) in the operator. Specifically, we consider the ensemble of Pauli operators supported inside a segment of $\ell$ sites in a 1D chain, in which each of the $\ell$ sites has a probability $\rho\in[0,1]$ of hosting a traceless Pauli (i.e., the average weight is $\rho\ell$). The sample complexity of learning a {\it typical} operator\footnote{Here for the ``typical'' value of a random variable $x$ we use its geometric mean $e^{\mathbb{E} \ln(x)}$. This quantity is less affected by rare fluctuations of $x$ compared to the mean $\mathbb{E}(x)$. This is important in our case since the shadow norm $\shadownorm{P}^2$ fluctuates across many orders of magnitude.} from this ensemble is given by
\begin{align}
%{\rm typ}(\| P \|_{\rm sh}^2) 
e^{\mathbb{E}_P \log \shadownorm{P}^2} 
& = [\lambda_{\bullet\bullet}^{-\rho^2/2} \lambda_{\bullet\circ}^{-\rho(1-\rho)}]^\ell . \label{eq:shdn_typ}
\end{align}
This is to be compared with the analogous result for random Pauli shadows, $3^{\rho \ell}$. Entanglement in the measurement basis is found to be beneficial only above a threshold Pauli density,
\begin{equation}
\rho > \rho^\ast(\delta) = \left[ 1 - \frac{\log(5-2e^\delta)}{2\log(e^\delta-1)} \right]^{-1},
\end{equation}
that depends on the amount of entanglement in the basis (parametrized by $\delta = \ln(2)-S_2^a$).
We find that $\rho^\ast(\delta)$ approaches 1 for $\delta\to0$ (Bell shadows) and 1/2 for $\delta\to\ln(2)$ (Pauli shadows). Thus in particular for any density $\rho \geq 1/2$, there exist entangled bases that are advantageous over product bases for this task.
We can also compare Eq.~\eqref{eq:shdn_typ} to the analogous result for shallow shadows at depth $\log(\ell)$, which is\footnote{For finite $\rho$, typical operators contain holes whose size is at most $O(\log \ell)$: the probability of a hole of length $x$ goes as $\sim (1-\rho)^x$, so the probability of having no such hole is roughly $\sim [1-(1-\rho)^x]^\ell$, which becomes large when $x \sim \log \ell$. In the limit of large $\ell$, at depth $O(\log \ell)$ such holes are filled and the interior of the operator is nearly equilibrated, giving the same result we would get for an operator with contiguous support ($\rho=1$).} $2^\ell$~\cite{ippoliti_operator_2023}; numerical inspection shows that locally-entangled shadows remain advantageous over shallow shadows above a threshold density $\rho^\ast \gtrsim 0.945$. 
Thus the results in Fig.~\ref{fig:qBshadows}, about 1D string operators with contiguous support, are qualitatively robust to the insertion of sufficiently sparse holes in the support.

\section{Beyond two-qubit measurements} \label{sec:ghz}

It is straightforward to generalize the previous discussion to measurements that factor into $n$-qubit bases, with $n>2$. 
For $n=3$, this requires picking a trimer covering and a choice of measurements on each trimer. It is easy to show\footnote{Eq.~\eqref{eq:lambda_A} for any pure measurement basis gives $\lambda_{\bullet\bullet\bullet} = (7-6\overline{\mathcal{P}})/27$ where $\overline{\mathcal{P}}$ is the average purity of a single qubit in the measurement basis (averaged over sites and states in the basis). This is maximized for $\overline{\mathcal{P}} = 1/2$, which is achieved by the GHZ basis.} that the optimal protocol (optimized for the learnability of compatible operators, defined here as having either 0 or 3 non-identity operators per trimer) is given by measurements in locally-scrambled GHZ bases, i.e., the 8 orthonormal states stabilized by $\pm X_1 X_2 X_3$, $\pm Z_1 Z_2$, $\pm Z_2Z_3$, up to local Clifford transformations.

While the optimization is less straightforward for $n>3$, we show next that the $n$-qubit GHZ basis is optimal among stabilizer measurements (i.e., measurements of commuting Pauli operators) at large $n$.
The relevant eigenvalue for compatible operators, adapting Eq.~\eqref{eq:lambda_A}, is
\begin{equation}
    \lambda_{[n]} = (-1/3)^n \sum_{A \subseteq [n]} (-2)^{|A|} \overline{\mathcal P}_A,
\end{equation}
where $[n] \equiv \{1,\dots n\}$.
The entanglement feature of an $n$-qubit GHZ state is $\overline{\mathcal P}_A = 1/2$ for all subsystems $A$ except $A = \emptyset$ and $A = [n]$, where $\overline{\mathcal P}_A = 1$; thus 
\begin{align}
    \lambda_{[n]}
    & = (-1/3)^n \left[ 1 + (-2)^n + \sum_{l = 1}^{n-1} \binom{n}{l} (-2)^l \frac{1}{2} \right] \nonumber \\
    & = \frac{2^n + 1 + (-1)^n}{2 \cdot 3^n}.
\end{align}
Thus the shadow norm of Pauli operators that are compatible with the partition is $\shadownorm{P}^2 = (f_n)^k$, with $k$ the Pauli weight and 
\begin{equation}
    f_n = 
    \left\{ 
    \begin{aligned}
    & \frac{3}{[2^{n-1}+1]^{1/n}} & & \text{if $n$ is even,} \\
    & \frac{3}{2^{1-1/n}} & & \text{if $n$ is odd.}
    \end{aligned}
    \right. \label{eq:ghz_scaling}
\end{equation}
Let us unpack this result:
\begin{itemize}
    \item For $n=1$ we recover random Pauli shadows, $f_1 = 3$;
    \item For $n=2$ we recover Bell shadows, $f_2 = \sqrt{3} \simeq 1.732$;
    \item For $n=3$ we find $f_3 = 3 / 2^{2/3} \simeq 1.890$, larger than $f_2$ and thus less efficient than Bell shadows when both are applicable, but potentially useful to learn quantities such as multi-point functions of 3-body operators;
    \item For $n\geq 4$, $f_n$ decreases monotonically and asymptotes to $3/2$, meaning a scaling of shadow norms as $\shadownorm{P}^2 \to (3/2)^k$ for compatible operators in the $n\gg 1$ limit.
\end{itemize}

It is easy to see that the latter scaling of $(3/2)^k$ is optimal for stabilizer measurements, i.e. when each run of the experiment measures $n$ independent, commuting Pauli operators $\{g_i\}_{i=1}^n$.
In effect this corresponds to measuring $2^n$ stabilizers: $\{ \prod_i g_i^{s_i}:\ \mathbf s\in \{0,1\}^n \}$. Thus each basis choice ``hits'' at most $2^n$ out of the $3^n$ maximum-weight Pauli operators. It follows that the probability of ``hitting'' a given compatible operator is $\leq [(2/3)^n]^{k/n} = (2/3)^k$, hence the bound on the shadow norm $\shadownorm{P}^2 \geq (3/2)^k$ for all stabilizer measurement bases. 
%It would be interesting to investigate optimality of this scaling more generally, beyond stabilizer measurements.

By the above reasoning, minimizing the sample complexity of learning ``compatible'' operators corresponds to choosing a stabilizer group in which the largest possible number of elements have full support. The distribution of weights across elements of a stabilizer group is a well-studied object with applications in coding theory, known as the Shor-Laflamme distribution and characterizable algebraically by {\it weight enumerator polynomials}~\cite{shor_quantum_1997,cao_quantum_2023}. The GHZ state is known to maximize the number of fully-supported elements in its stabilizer group~\cite{miller_shor-laflamme_2023}, meaning that it is indeed optimal for our task. 

The improved scaling of sample complexity in these protocols with larger $n$ however comes with some trade-offs. 
For one, the set of ``compatible'' operators whose shadow norm obeys the scaling in Eq.~\eqref{eq:ghz_scaling} gets more constrained with increasing $n$ (the set comprises $[3^n+1]^{N/n}$ operators). 
Secondly, preparing the GHZ measurement basis on $n$ qubits requires either circuit depth linear in $n$, which makes the method less scalable on noisy hardware and limits $n$ to modest finite values, or the introduction of extensively many auxiliary qubits. The performance of these schemes under realistic constraints is an interesting direction for future work.

\section{Discussion}\label{sec:discussion}

\subsection{Summary}

We have introduced classical shadows protocols based on randomized measurements that feature entanglement over finite-sized subsystems. These protocols are NISQ-friendly, requiring only local shallow circuits; further, the classical post-processing steps are straightforward, on par with the standard classical shadows protocols based on random Pauli or Clifford measurements~\cite{huang_predicting_2020}.

\begin{table*}
\centering
\begin{tabular}{lccc} \toprule
Measurement	& Sample complexity & Number of & Fraction of  \\ 
protocol	& scaling & compatible Paulis & compatible Paulis \\
\midrule
Pauli & $3^k$ & all ($4^N$) & 1 \\
General 2-qubit & Eq.~\eqref{eq:shdn_delta} & all ($4^N$) & 1 \\ 
Bell & $3^{k/2} \simeq 1.732^k$ & $10^{N/2}$ & $\simeq 0.791^N$ \\ 
%(3/2^{2/3})^k 
GHZ ($n=3$) & $\left( \frac{3}{2^{2/3}} \right)^k \simeq 1.890^k$ & $28^{N/3}$ & $\simeq 0.759^N$ \\
GHZ (general $n$) & Eq.~\eqref{eq:ghz_scaling} & $[3^n+1]^{N/n}$ & $\left[ \frac{3}{4}(1+3^{-n})^n \right]^N$ \\ \bottomrule
\end{tabular}
\caption{Summary of results for various locally-entangled classical shadows protocols. \label{tab:summary}}
\end{table*}

We have shown that locally-entangled measurements can lead to substantial improvements in sample complexity for some Pauli estimation tasks.
As a paradigmatic example, we have focused on two-qubit Bell measurements, Sec.~\ref{sec:bell}. These achieve an improved scaling of sample complexity $\propto 3^{k/2}$ with Pauli weight $k$ for many operators, while failing to learn others. Such trade-offs are unavoidable since the scaling $\sim 3^k$ of random Pauli shadows is optimal in general under a fixed set of local measurements~\cite{huang_predicting_2020} (i.e. it is not possible to do better for all Pauli operators);
nonetheless, the trade-off can be advantageous for tasks of interest in quantum many-body physics such as the estimation of string operators or multi-point functions of local operators.

We have further shown (Sec.~\ref{sec:quasi-bell}) that general two-qubit entangled bases make it possible to interpolate between Pauli and Bell measurements. This gives a family of tomographically-complete protocols that retain a favorable scaling of sample complexity in many cases, Eq.~\eqref{eq:shdn_delta_approx}. 
Finally, when allowing entanglement between $n>2$ qubits (Sec.~\ref{sec:ghz}), we have found that the optimal basis is given by $n$-qubit GHZ states, which enable the estimation of certain `compatible' Pauli operators with even more advantageous scaling of the shadow norm, $\shadownorm{P}^2\sim (3/2)^k$ for large $n$, Eq.~\eqref{eq:ghz_scaling}. 
Our results are summarized in Table~\ref{tab:summary}.

\subsection{Connections with prior work}\label{sec:previous}

\subsubsection{Shallow shadows}

Recent works have focused on leveraging locality---an important factor in many NISQ architectures---to develop variations of classical shadows with certain practical advantages.
In particular, {shallow shadows}~\cite{akhtar_scalable_2023,bertoni_shallow_2022} implement the basis randomization step via circuits of variable depth and have been recently shown to give a significantly improved sample complexity (relative to random Pauli measurements) for learning expectation values of geometrically local Pauli operators, among other advantageous properties~\cite{akhtar_scalable_2023,bertoni_shallow_2022,arienzo_closed-form_2022, ippoliti_operator_2023}.

Bell shadows (and other $n=2$ protocols) may be seen as a depth-1 instance of shallow shadows, featuring a single layer of two-qubit entangling gates.
However, this appears to raise a puzzle in relation to the shadow norm of Pauli string operators in 1D.
While shallow shadows were shown to achieve an optimal scaling of sample complexity $\sim k2^k$~\cite{ippoliti_operator_2023}, Bell shadows manage to achieve the improved scaling $\sim 3^{k/2}$ for some of the same operators. How can a depth-1 instance outperform the optimal-depth version of the protocol?

The answer lies in the fact that Bell shadows feature a partly deterministic measurement sequence: 
the state is locally-scrambled, then evolved by a deterministic circuit (a single layer of entangling $\CZ$ or $\CPhase{\phi}$ gates) and measured in a deterministic local basis ($X$). 
Shallow shadows, on the contrary, are based on fully-random circuits~\cite{bertoni_shallow_2022,arienzo_closed-form_2022}, or feature a local scrambling step right before the final single-qubit measurements~\cite{hu_classical_2023,akhtar_scalable_2023}.
Exploiting this final local-scrambling step, Ref.~\cite{ippoliti_operator_2023} maps the shadow norm to properties of the operator weight distribution under the twirling circuit, and as a consequence derives the optimal scaling\footnote{This is the same scaling one would obtain from a tensor product of random Clifford bases on $k$-qubit segments ($\sim 2^k$), iterated $k$ times to address all possible Pauli endpoint locations modulo $k$. Surprisingly, shallow shadows achieve this scaling already at depth $O(\ln k)$~\cite{ippoliti_operator_2023}.} $\sim k 2^k$.
Bell shadows, by avoiding the local scrambling step before the final single-qubit measurements, evade this result and are thus able to improve the scaling to $\sim 3^{k/2}$ in the best case. 

Another advantage of locally-entangled shadows over generic shallow shadows is that the inversion of $\mathcal{M}$ is effortless (on par with the random Pauli and random Clifford protocols~\cite{huang_predicting_2020}), and does not require tensor network algorithms~\cite{akhtar_scalable_2023,bertoni_shallow_2022}. This straightforwardly unlocks applications to higher-dimensional systems. 
Moreover, locally-entangled shadows may also be more advantageous for learning $k$-local, but geometrically non-local Pauli operators (as long as these are compatible with the partition of the system into $n$-qubit sets)---see the discussion on multi-point correlators in Sec.~\ref{sec:bell_app}. 

At the same time, generic shallow shadows retain some significant advantages. First of all, shallow shadows can efficiently learn many-body fidelity (at depth $t \sim \log(N)$), akin to random Clifford measurements~\cite{bertoni_shallow_2022}. This is beyond the scope of locally-entangled shadows. 
Within Pauli estimation, generic shallow shadows may be more efficient for estimating expectation values of local Pauli operators in 1D with a finite density of ``holes'' (identity operators) between their endpoints, depending on the density of holes---see the discussion at the end of Sec.~\ref{sec:quasi-bell}.

\subsubsection{Learning via locally-entangled measurements}

Various other state-learning protocols that make use of Bell, GHZ, or other locally-entangled measurements have been studied.
Refs.~\cite{hamamura_efficient_2020,kondo_computationally_2022,escudero_hardware-efficient_2023} focus on including Bell and GHZ measurements as part of a measurement optimization algorithm for learning a given set of operators, which is complementary to our focus on randomized measurements. 

Ref.~\cite{jiang_optimal_2020}, while written prior to the introduction of classical shadows, has several analogies with our discussion of Bell shadows, Sec.~\ref{sec:bell}. 
The approach relies on introducing one auxiliary qubit $\vec{\tau}_i$ per system qubit $\vec{\sigma}_i$; each of the auxiliary qubits is in an initial state $\xi = (\mathbb{I} + \hat{n}\cdot \vec{\tau})/2$ with $\hat{n} = (1,1,1)/\sqrt{3}$. By measuring the two qubits $\vec{\sigma}_i$, $\vec{\tau}_i$ in the standard Bell basis, one simultaneously learns the expectation all three commuting operators\footnote{A similar idea was recently employed to embed non-commuting Hamiltonians into commuting ``Ising'' models in an enlarged space~\cite{verresen_everything_2023}.} $\sigma^\alpha_i \otimes \tau^\alpha_i $ for all $\alpha = x,y,z$ and each site $i$.
We thus learn the expectation of $\tilde{P} = \bigotimes_i \sigma_i^{\alpha_i} \otimes \tau_i^{\alpha_i}$: 
\begin{align}
    \langle \tilde{P} \rangle 
    & = \langle \bigotimes_i \sigma_i^{\alpha_i} \otimes \tau_i^{\alpha_i} \rangle = \langle P\rangle \prod_i \Tr(\xi_i \tau_i^{\alpha_i}) \nonumber \\
    & = 3^{-k/2} \langle P\rangle,
\end{align}
where $P = \bigotimes_i \sigma_i^{\alpha_i}$ is a Pauli operator on the system qubits, and $k$ is its weight. 
Learning $\langle P \rangle$ with additive error $\epsilon$ requires learning $\langle \tilde{P}\rangle$ with additive error $3^{-k/2}\epsilon$, thus the sample complexity scales as $3^k \epsilon^{-2}$.

The relationship with Bell shadows becomes apparent if one views the above protocol as learning a Pauli operator $\tilde{P}$ of weight $2k$ on a two-leg ladder. The dimer covering is given by pairing qubits $\vec{\sigma}_i$ and $\vec{\tau}_i$ on each rung, and $\tilde{P}$ is manifestly compatible with the covering, giving the expected sample complexity $\sim 3^{|\tilde{P}|/2} = 3^k$. 
Thus the protocol of Ref.~\cite{jiang_optimal_2020} may be seen as a ``derandomized'' version of Bell shadows with a specific geometry, dimer covering, and subset of initial states.

\subsection{Future directions}\label{sec:future_science}

The advantages and limitations of locally-entangled shadows both follow from using a structured, (partially) non-random circuit. 
This observation points to interesting directions for future research, based on leveraging structure in a problem of interest to design tailored, highly-optimized shadow (or shadow-like) state-learning protocols. 
Several such proposals have been put forth, e.g. with the goal of optimizing energy estimates for molecular Hamiltonians~\cite{hadfield_measurements_2020,huang_efficient_2021,hadfield_adaptive_2021,hillmich_decision_2021,yen_deterministic_2023,wu_overlapped_2023}. However, these proposals typically deal with the optimization of local basis choices within a Pauli measurement framework. Our work points to the possibility of substantial gains from using entangled measurement bases, even in the simplest and most practically accessible case of two-qubit entanglement.
It would be interesting to identify more complex many-body entanglement structures optimized for learning different classes of properties of quantum states, such as entropies or other nonlinear functionals.

Another interesting direction is to bring locally-entangled shadows within reach of analog quantum simulators, by adapting the approaches of Refs.~\cite{tran_measuring_2023, mcginley_shadow_2022}. 
{\it In lieu} of random unitary gates, these approaches make use of fixed Hamiltonian dynamics and the stochastic nature of quantum measurements to supply the randomization needed for classical shadows without digital control~\cite{choi_preparing_2023,cotler_emergent_2023,ho_exact_2022,claeys_emergent_2022,ippoliti_dynamical_2023,ippoliti_solvable_2022, claeys_universality_2023}.
The ``patched quench'' scenario of Ref.~\cite{tran_measuring_2023} in particular appears as a natural setup for optimized locally-entangled shadows. The task of finely tuning the amount of entanglement in the measurement basis for these protocols is nontrivial, but may be achievable depending on the nature of the analog simulator dynamics.

\acknowledgments
I thank Vedika Khemani, Yaodong Li and Tibor Rakovszky for collaborations on related topics, and Daniel Miller for bring the results of Ref.~\cite{miller_shor-laflamme_2023} to my attention.
This work is supported  in  part  by  the Gordon and Betty Moore Foundation's EPiQS Initiative through Grant GBMF8686. 

\bibliographystyle{quantum}
\bibliography{le_shadows}

\end{document}